\documentclass[preprint,11pt]{aastex}

 \newcommand{\degree}{\mbox{$^{\circ}$}}
 \newcommand{\am}{\mbox{\arcmin}}
 \newcommand{\as}{\mbox{\arcsec}}

 \newcommand{\kms}{\mbox{km s$^{-1}$}}

 \newcommand\submm{submillimeter}

 \newcommand{\lsun}{\mbox{L$_\odot$}}
 \newcommand{\msun}{\mbox{M$_\odot$}}

 \newcommand{\td}{\mbox{$T_d$}}
 \newcommand{\lbol}{\mbox{$L_{bol}$}} 
 \newcommand{\tbol}{\mbox{$T_{bol}$}} 

 \newcommand{\rinf}{\mbox{$r_{inf}$}} 
 \newcommand{\fsmm}{\mbox{$L_{smm}/L_{bol}$}} 
 \newcommand{\lsmm}{\mbox{$L_{smm}$}} 

 \newcommand{\hh}{\mbox{{\rm H}$_2$}}
 \newcommand{\form}{H$_2$CO}

 \newcommand{\cooo}{C$^{18}$O}
 \newcommand{\coooo}{C$^{17}$O}
 \newcommand{\hcop}{HCO$^+$}
 \newcommand{\hcopi}{H$^{13}$CO$^+$}
 \newcommand{\dcop}{DCO$^+$}
 \newcommand{\nthp}{N$_2$H$^+$}
 \newcommand{\ntdp}{N$_2$D$^+$}

 \input{epsf}

\begin{document}


 \title {\bf The Spitzer c2d Survey of Nearby Dense Cores VII: Chemistry and Dynamics in L43}
 \author {Jo-Hsin Chen \& Neal J. Evans II}
 \affil{Department of Astronomy, The University of Texas at Austin,
        Austin, Texas 78712--1083}
 \email{jhchen@astro.as.utexas.edu, nje@astro.as.utexas.edu}
 \author{Jeong-Eun Lee}
 \affil{Department of Astronomy and Space Science, Astrophysical Research Center for the Structure and Evolution of the Cosmos, Sejong University, Seoul 143-747, Korea}
 \email{jelee@sejong.ac.kr}
\and
 \author{Tyler L. Bourke}
 \affil{Harvard-Smithonian Center for Astrophysics, 60 Garden St., Cambridge MA 02138, USA}
 \email{tbourke@cfa.harvard.edu}                 
 \begin{abstract}
We present results from the Spitzer Space Telescope and molecular line observations of 9 species toward the dark cloud 
L43. The Spitzer images and molecular line maps suggest it has a starless core and a Class I protostar evolving in the 
same environment. CO depletion is seen in both sources, and {\dcop} lines are stronger toward the starless core. With a goal of testing the chemical characteristics from pre- to protostellar stages, we adopt an 
evolutionary chemical model to calculate the molecular abundances and compare with our observations. Among the 
different model parameters we tested, the best-fit model suggests a longer total timescale at the pre-protostellar stage, 
but with faster evolution at the later steps with higher densities. 

 \end{abstract}


 \section{Introduction}
The first step in our understanding of star formation is to determine the properties of dense cores through observations. In general, dust continuum emission is useful to study physical structures of the envelope with tools of radiative transfer modeling. Molecular emission lines are complementary to trace physical properties through studies of excitation; also the line profiles can uniquely probe kinematics. However, the molecular abundances are not homogeneous for all species throughout the core, which easily leads to confusion when interpreting the observed lines. For example, the effect of selective depletion is observed: carbon-bearing and sulfur-bearing molecules are easily depleted in the inner cold, dense region of starless cores, and nitrogen-bearing molecules can remain in the gas phase at higher densities (e.g. Caselli et al. 1999; Tafalla et al. 2002, 2004; Lee et al. 2003). Once a central object forms, the chemistry becomes more complicated since the temperature distribution can be drastically affected by the additional heating mechanisms such as radiation of the central protostellar object and shocks from outflow. The surrounding materials are heated up and reach the evaporation temperature of CO, the main gas-phase repository of carbon. The abundances of other molecules such as {\hcop} and {\form} can be enhanced by their desorption off dust grains as well as the higher formation rates due to the increase of gas-phase CO abundance. The desorbed CO can also destroy {\nthp}, resulting in a drop in {\nthp} abundance, and an anti-correlation of {\cooo} and {\nthp} emission has been seen toward a few Class 0 protostars (e.g. J{\o}rgensen 2004a; Lee et al. 2007) on $\sim10^3$ AU scale. Such chemical characteristics obtained from molecular spectra can potentially be used to constrain the evolutionary state, but a comprehensive picture of coupled dynamical and chemical evolution is needed since the chemistry depends on not only elapsed time but also density and temperature. Moreover, our knowledge of critical chemical processes, which needs laboratory experiments relevant to interstellar environments, is still limited. 

L43 is a Lynds opacity class 6 (i.e. the highest opacity) dark nebula (Lynds 1962) located near the Ophiuchus cloud at a distance of 130 pc (de Geus et al. 1989). A bipolar molecular outflow driven by IRAS 16316-1540 (Andr\'{e} \& Montmerle 1994) is seen, as well as a circumstellar envelope elongated perpendicular to the outflow axis (Mathieu et al. 1988; Lee et al. 2002; Lee \& Ho 2005). The northern outflow lobe is redshifted, and the southern outflow lobe is blueshifted. A reflection nebula RNO91 is associated with the outflow source (Hodapp 1994). Two sources (hereafter RNO91 and L43E) are detected at submillimeter wavelengths (e.g., Ward-Thompson et al. 1999; Shirley et al. 2000; Kirk et al. 2005; Young et al. 2006). It is usually assumed that RNO91 is equivalent to IRAS 16316-1540, and L43E is a starless core. The mass within 3-${\sigma}$ contours is estimated to be $0.3\pm0.1$ {\msun} for the dense core associated with RNO91 and $0.8\pm0.2$ {\msun} for L43E from the Submillimeter Common-User Bolometer Array (SCUBA) 850 {\micron} observations (Young et al. 2006). The two sources are roughly 80{\as} apart and provide a test bed for chemical signatures prior to and during collapse. 

A few starless cores and cores with central protostars have been previously studied in detail with a combination of modeling and observations. Typically, envelope density and temperature distributions are determined with dust radiative transfer modeling, which provides the inputs for line simulation, and the radial dependence of molecular abundances can be tested. For example, empirical abundance models, such as a step function, are used to describe the amount of depletion in starless cores (e.g. Lee et al. 2003). For sources with central heating, a drop function with the innermost region undepleted can reasonably reproduce the observed lines (e.g. J{\o}rgensen et al. 2004b). Abundance profiles predicted by evolutionary chemical models with varying densities and temperature are also considered; for example, Aikawa et al. (2005) examined the observed properties of 4 prestellar cores with contracting Bonnor-Ebert (BE) spheres, and Evans et al. (2005) presented line modeling of a collapsing protostar, B335, with a chemical model incorporating inside-out collapse (Shu 1977), described in Lee et al. (2004). In this paper, we also aim at testing such dynamical and chemical models, but L43 provides two snapshots, with and without protostellar heating, which allow us to compare the chemical behavior at different evolutionary stages based on the same initial environmental condition. 

This paper presents new observations of molecular spectra and mid-infrared observations toward L43, and we follow the method of Lee et al. (2004) in attempts to interpret the data, including a dynamical model, continuum radiative transfer, gas energetics, and line radiative transfer. The chemical network is updated with a new binding energy of N$_2$ (\"{O}berg et al. 2005), and we used the {\nthp} dissociative recombination rates of Geppert et al. (2004). We have also extended the network to include deuterated species since deuterium fractionation is important. This is especially true in early phases when it is associated with CO depletion. \S 2 describes the observations, \S 3 presents observational results and simple analyses, \S 4 details the modeling sequence, and the modeling results are discussed in \S 5.

 \section{Observations}
\subsection{Spitzer Space Telescope}
Mid-infrared observations of L43 were performed with the Spitzer Infrared Array Camera (IRAC; Fazio et al. 2004) and Multiband Imaging Photometer for Spitzer (MIPS; Rieke et al. 2004) aboard the Spitzer Space Telescope by the Spitzer legacy Program, From Molecular Cores to Planet-Forming Disks (c2d; Evans et al. 2003). The IRAC observations (Astronomical Observation Request [AOR] ID 0005121280, 0005121792) were obtained in all 4 bands (3.6, 4.5, 5.8, and 8.0 {\micron}) on 2004 August 16 and 17. Two epochs of observation were taken to allow the identification and removal of the asteroids, each with 10{\as} dithering, a 12 second exposure, and a 0.6 second of the High-Dynamic Range (HDR) mode image. The MIPS observations (AOR ID 0009409792, 0009433344) were also taken in 2 epochs at the first 2 bands, 24 {\micron} and 70 {\micron}, on 2004 March 10. We integrated 1 cycle of 3 seconds for 24 {\micron}, and 3 cycles of 3 seconds for 70 {\micron}.

The data were processed with pipeline version S13.2.0 of Spitzer Science Center (SSC), and improved by the c2d team with corrections of artifacts. Detailed discussions of the data processing and products by the c2d team is included in the delivery documentation (Evans et al. 2007), as well as in Harvey et al. (2006) on IRAC data and Young et al. (2005) on MIPS data.

Followup observations that are significantly deeper than those of c2d were performed as part of Spitzer program 20386 (PI: Philip C. Myers).  The observations with IRAC were made on 2005 August 22 (AOR key 14612480), and with MIPS at 24 {\micron} only on 2006 April 8 (AOR key 14617600).  For the IRAC observations the on-source time of 480 seconds (16 dithers of 30 seconds each) is ten times longer that used for c2d.  For the MIPS 24 {\micron} observations the on-source time is 333 seconds, or about 7 times longer than c2d.

Followup observations with MIPS at 24, 70 and 160 {\micron} were also performed, as part of Spitzer program 30384 (PI: Tyler L. Bourke), on 2007 April 14 (AOR key 18158336).  The observations utilized scan map mode in all three bands. The data reductions were carried out as described in Stutz et al.(2007). The MIPS images are 15{\farcs}5 $\times$ 54{\arcsec} in size, with exposure times of $\sim$150 seconds at 24 {\micron}, $\sim$90 seconds at 70 {\micron}, and $\sim$25 seconds at 160 {\micron} (these are approximate as the sampling is non-uniform).  The resolutions of the observations are $\sim6{\arcsec}$, $\sim18{\arcsec}$, and $\sim40{\arcsec}$ at 24, 70 and 160 {\micron}, respectively.

\subsection{Molecular Line Observations}
Line observations were carried out with the CSO\footnote{The Caltech Submillimeter Observatory is supported by the NSF} in the period 1998 July to 2006 June. We used a double-side band SIS receiver with an acousto-optical spectrometer (AOS) with a bandwidth of 50 MHz over 1024 channels. The pointing uncertainty was about 3.6{\as} on average. The observed lines with the reference frequency, the date of observation, the full width at half maximum beam size ($\theta_b$), the main beam efficiency ($\eta_{mb}$), and the velocity resolution ($\delta$v) are summarized in Table 1. Assuming the source coupling efficiency is unity because the sources are extended, we can calculate the radiation (source) temperature ($T_R$) given by $T_R = T^\ast_A/\eta_{mb}$, where $T^\ast_A$ is the antenna temperature corrected for atmospheric attenuation. 

Based on the {\submm} continuum emission mapped with SCUBA (Shirley et al. 2000) at 450 and 850 {\micron}, we use the  (0{\as}, 0{\as}) position of (16$^h$34$^m$29.0$^s$, $-$15\degree47\am3.2\as) in J2000.0 coordinates, and assume that the centroids of RNO91 and L43E are at (7{\as}, 5{\as}) and (89{\as}, 6{\as}) respectively. The observations were performed with position switching at the off position (900{\as}, 0{\as}). All spectra were then reduced with the analysis routines of the package CLASS. Some of the data were taken with different offsets, and in cases with a map, we have synthesized the spectra to the centroid positions at a risk of slight degradation of the spatial resolution.

 \section{Results}
 \subsection{Source Properties}
Figure 1 shows a 3-color image of L43 in the IRAC bands 1, 2, and 4. Figure 2 shows the 850 {\micron} contours from Young et al. (2006) on IRAC and MIPS images at the 3.6, 8.0, 70, and 160 {\micron} bands, as well as a 350 {\micron} map from Wu et al. (2007) obtained with the Submillimeter High Angular Resolution Camera (SHARCII) at the CSO. L43E is seen as a dark core at 8.0 {\micron} and is traced well by the 850 {\micron} dust emission. Photometry including the Spitzer results, the submillimeter results, and other observations are listed in Table 2. The table gives the wavelength ($\lambda$), flux density ($S_\nu$) with uncertainty ($\sigma$), and the aperture size. In cases of no detection, an upper limit is given.  

RNO91 falls in the range of Class I sources, as described in Allen et al. (2004) with the IRAC colors $[3.6]-[4.5] = 0.93$, $[5.8]-[8.0] = 0.79$. For L43E, there is no point source detected in IRAC or MIPS bands. From all data of RNO91 in Table 2, we also calculate the bolometric luminosity ({\lbol}), the bolometric temperature ({\tbol}), and {\fsmm}, where {\lsmm} includes flux at ${\lambda} > 350 {\micron}$. The {\lbol} is $2.5\pm0.1$ {\lsun}, and {\tbol} is $337.6\pm18.9$ K, which fits the criterion for Class I in Chen et al. (1995). With {\fsmm} = 0.004, RNO91 can also be categorized as Class I as defined by Andr\'{e} et al. (1993).

 \subsection{Molecular Lines}
Spectra toward the centroid position of both sources are presented in Figure 3. The line parameters derived from the spectra are presented in Table 3, which gives the integrated intensity (I), the velocity with respect to the local standard of rest (v$_{LSR}$), the linewidth ($\Delta$v), $T_R$, and the number of observed positions. For single-peaked lines, these values are determined from a Gaussian fit. For double peaked-lines, $T_R$ is from the stronger peak, v$_{LSR}$ is the velocity of the dip, both determined by eye, I is integrated over the full line, and the $\Delta$v is I divided by $T_R$. For lines with hyperfine structure, v$_{LSR}$ and $\Delta$v are from hyperfine structure fitting, I is the total area under these components, and $T_R$ is for the strongest component.

The centroid velocities are estimated to be $0.52\pm0.01$ {\kms} for RNO91 and $0.77\pm0.02$ {\kms} for L43E from the single peaked lines {\dcop} and {\cooo}, which are likely to be optically thin. Figure 4 shows the {\dcop} map in the center region of L43. The velocities are blueshifted toward the east and redshifted toward the west, possibly indicating rotation of the cloud that encompasses both sources. The {\dcop} lines are slightly wider near the RNO91 peak, but no significant variations in linewidths are seen across the mapped region. Figure 5 shows the integrated intensity map of {\coooo} and {\cooo}; neither of the two CO isotopologues shows line peaks near the center position of RNO91, which could result from CO depletion, suggesting that the evaporation from central heating is still small. Figure 6 shows the 350 {\micron} contours (Wu et al. 2007) overlaid on the extended {\dcop} integrated intensity map. Several {\submm} peaks are resolved in the 350 {\micron} map, and they lie on the broad extended structure seen in {\dcop}, suggesting that they formed from the same structure. CS $J = 2-1$ was mapped by Mathieu et al. (1988) and Lee et al. (2005) for RNO91 tracing both the outflow and the envelope, but the CS $J = 7-6$ line is not detected in our observations.

\subsection{Simple Analysis}
\subsubsection{The {\hh} column density from dust continuum emission}
Assuming the dust grains emit as blackbodies ($I_{\nu}= {\kappa}_{\nu}B_{\nu}(T_D)$, $B_{\nu}$ is the Planck function), and the emission is optically thin at a submillimeter wavelength, we can relate the observed flux density ($S_{\nu}$) to the column density of gas ($N$(H$_2$)) by
$$N({\hh})=\frac{S_{\nu}}{{\mu}m_{H}{\kappa}_{\nu}B_{\nu}(T_{D})\Omega},$$
where $\mu$ is the mean molecular weight, $m_{H}$ is the atomic mass, ${\kappa}_{\nu}$ is the dust opacity per gram of gas, and $\Omega$ is the aperture solid angle.
We take the 850 {\micron} data in Table 2 and measure the flux density in an aperture of 34{\arcsec} to match the main beam size of {\cooo} observations. We adopt the dust opacities from column (5) of Table 1 in Ossenkopf \& Henning (1994, hereafter OH5) for coagulated dust grains with thin ice mantles, and assume a standard gas-to-dust ratio of 100 to get ${\kappa}_{850}=0.02$ cm$^2$ g$^{-1}$. Assuming {\td} of 10K, we can obtain 
$$N({\hh})_{RNO91,{S_{850}}}= 3.35 \times 10^{22} cm^{-2}; $$
$$N({\hh})_{L43E,S_{850}}= 4.76 \times 10^{22} cm^{-2}.$$
If {\td} is 20K, the calculated N({\hh}) would decrease by a factor of 3. If we assume a gas-to-dust mass ratio of $\sim$124, for example, using the model from Table 4 of Draine (2003; and the references therein), the derived $N(H_2)$ would be 24\% higher.

\subsubsection{The {\hh} column density from molecular lines}
Assuming the line is optically thin and all levels in LTE, the integrated intensity for the transition $J {\rightarrow} J - 1$ of a linear molecule $x$ in $^1{\Sigma}$ state and the column density ($N(x)$) of species $x$ is given by
$$N(x) = {\frac{3kQe^{\frac{E_J}{kT_{ex}}}}{8{\pi}^3{\nu}{\mu}^2J}}\int T_R\,dv$$ 
as in Equation (2) from Lee et al. (2003), where $E_J = hBJ(J + 1)$ in the approximation of rigid-rotor, B is the rotational constant, $\mu$ is the dipole moment, Q is the partition function, ${T_{ex}}$ is the excitation temperature, and $T_R$ is the measured brightness temperature. We can correct this equation for optical depth:
$$N(x)_{thick} = N(x)_{thin}\frac{{\tau}_{\nu}}{1 - e^{-{\tau}_{\nu}}}.$$  
We can also convert the column density of a certain molecule into {\hh} column density by
$$N({\hh}) = \frac{N(x)}{X(x)},$$
where $X(x)$ is the molecular abundance.
Under the assumption that all lines are thermalized and $T_{ex} = 10 K$, we calculate $N({\hh})$ using the observed {\cooo} J = 2 - 1 line averaged over the beam (I in Table 3). We adopt the physical parameters B, $\mu$, and Q from the JPL catalog, $X$(CO) of $2.7 \times 10^{-4}$ (Lacy et al. 1994), and $X$(CO)/$X$({\cooo}) $\sim 580$ (Wilson \& Rood 1994). If {\cooo} is optically thin, we obtain
$$N({\hh})_{RNO91,C^{18}O}= 1.96 \times 10^{21} cm^{-2};$$
$$N({\hh})_{L43E,C^{18}O}= 1.74 \times 10^{21} cm^{-2}.$$

The results are less than one-fifteenth the values calculated from dust emission at 850 {\micron}. A lower X(CO), for example, X(CO) $\sim 1.8 \times 10^{-4}$ from Dickman (1978), can result in less CO depletion, but the ratio of $N$({\hh}) derived from {\cooo} to $N$({\hh}) derived from H$_2$ would be still less than 0.1 for both sources. Considering that {\cooo} J = 2 - 1 line might be optically thick, the {\hh} column densities derived from the line would be underestimated. To correct the optical depth effect, we can fit the {\coooo} J = 2 - 1 hyperfine structures to estimate the total optical depth of {\coooo} J = 2 - 1, and calculate the optical depth of {\cooo} J = 2 - 1 with the abundance ratio $X$({\cooo})/$X$({\coooo}) $\sim$ 4  \cite{wouterloot05}. The hyperfine fitting result of ${\tau}_{C^{17}O(2-1)} \sim 0.59$ for RNO91 has a large error due to the low S/N. After the correction for ${\tau}$, $N$({\hh}) derived from {\cooo} is still lower than $N$({\hh}) derived from H$_2$ by a factor of $\sim 10$, which strongly indicates the depletion of CO molecules. Furthermore, the column densities derived from {\cooo} are similar in both sources, but from the 850 {\micron} data, the column density is higher in L43E than in RNO91, which suggests there can be more depletion in L43E, with no central heating in the starless core. We can investigate the depletion in more detail through the modeling procedure in the next section where the distribution of excitation temperature and molecular abundances are taken into consideration.

\subsubsection{Deuterium Enhancement}
At low temperature, the production of H$_2$D$^+$ is favored through the exothermic exchange reaction H$_3^+$ + HD $\rightarrow$ H$_2$D$^+$ + H$_2$. The H$_2$D$^+$ abundance can be further enhanced with CO depletion, and its deuteration can propagate to other molecules, such as {\dcop} and {\ntdp}, by ion-molecule reactions (Dalgarno \& Lepp 1984). Therefore, in cold clouds, the molecular D/H ratio can be much higher than the elemental D/H ratio ($\cong1.5\times10^{-5}$; Oliveira et al. 2003).

The integrated intensities of {\dcop}(3-2) toward RNO91 and L43E are similar, but the intensity of {\hcop}(3-2) is much higher in RNO91 than in L43E, which must be due to the higher {\hcop} column densities and higher gas temperature in RNO91. We do not have optically thin lines such as {\hcopi} or HC$^{18}$O$^+$ to help determine the {\hcop} column densities (e.g. Appendix A1 in Caselli et al. [2002]), but we calculate $N$({\dcop}) and $N$({\hcop}) for the optically thin case (${\tau}_{DCO^+(3-2)}{\ll}1$ and ${\tau}_{HCO^+(3-2)}{\ll}1$). For constant abundances and optically thin lines, we obtained $N$({\dcop})$_{L43E}$/$N$({\hcop})$_{L43E}\sim4.11$ and $N$({\dcop})$_{RNO91}$/$N$({\hcop})$_{RNO91}\sim0.85$, assuming $T_{ex} = 10$ K. If $T_{ex}$ is 1 K lower ($T_{ex} = 9$ K), the column densities can increase by 25\% - 30\% for each molecule. The optically thin approximation gives a lower limit of 4.8 for the difference of the enhancement factor $N$({\dcop})/$N$({\hcop}) between L43E and RNO91.


In a CO-depleted region, {\ntdp} can be more enhanced than {\dcop} since it suffers from less depletion. Assuming the {\nthp} lines are optically thin, $N$({\nthp}) toward L43E can be derived to be $1.23\times10^{13} cm^{-2}$. Also, from \S 3.3.2, the CO depletion factor $(f_D)$ of L43E is $\sim22$. From the samples of starless cores in Crapsi et al. (2005), objects with higher {\nthp} column densities ($>10^{13} cm^{-2}$) and more CO depletion ($f_D>10$) tend to have higher $N$({\ntdp})/$N$({\nthp}) abundance ratio ($>0.1$) and can be recognized as the most evolved starless cores. However, the line intensity of {\nthp}(3-2) toward L43E is $\sim$3 times lower than RNO91 and {\ntdp} (3-2) is only marginally detected in L43E with a very narrow line. The $N$({\ntdp})/$N$({\nthp}) ratio we derived is $\sim$0.03. For RNO91, the $N$({\ntdp})/$N$({\nthp}) ratio we derived is $\sim 0.02$, which is at the lower end compared with the starless core samples because of heating from the central protostar.

\section{The Modeling Procedure}
In this section we describe the sequence of modeling from the determination of the dust and gas temperature distribution to line simulation. Similar techniques were employed in Young et al. (2004) and Evans et al. (2005), but in this paper, we apply an updated chemical network.

For a given density structure ($n(r)$), we can calculate $T_D(r)$ by computing the transport of energy through the dust envelope (\S 4.1), and we simulate the observed spectral energy distribution (SED) and the spatial intensity distribution ($I_{\nu}(b)$, where b is the impact parameter). Then $T_K(r)$ can be calculated from a gas energetics code (\S 4.2). We use an evolutionary model including the self-consistent treatment of dynamics, dust continuum radiative transfer, gas energetics, and chemistry (\S 4.3). Once we obtain the molecular abundances ($X(r)$), we can calculate the populations of relevant energy levels (n$_j(r)$), simulated line profiles, and compare the observations with the simulations (\S 4.4).

\subsection{Dust Models}
We use the radiative transfer package CSdust3 (Egan et al. 1988), which has input parameters of luminosity of the central source, dust opacity as a function of wavelength, the strength of the interstellar radiation field (ISRF), and the density distribution of the envelope. The program ObsSph (Shirley et al. 2002) then takes the output from CSdust3 and simulates the telescope observations. For simplicity and because of the limitation of one-dimensional model, we focus on the SED fitting at longer wavelengths to trace the envelope. At short wavelengths ($\lambda \leq 60 {\micron})$, the SED is more sensitive to opacity and detailed geometry. For example, RNO91 is known to have an inclined disk (Weintraub et al. 1994), and the infrared spectra also show inclination effects, such as silicate emission (e.g. Boogert et al. 2008). With an 1-dimensional model envelope, the amount of observed protostellar emission that directly comes through the outflow cavity is not taken into account, which can lead to an underestimate of fluxes, particularly at mid-IR wavelengths (e.g. Whitney et al. 2003a, 2003b; Robitaille et al. 2006; Crapsi et al. 2008). 

We use the ISRF spectrum based on Cosmic Background Explorer (COBE) data from Black (1994) with ultraviolet (UV) wavelengths described in Draine (1978), and modified by dust extinction with the properties in Draine \& Lee (1984) assuming an attenuation by an external A$_V$ outside the core we observe. A scaling factor s$_{ISRF}$ is also introduced to increase or decrease the ISRF at all wavelengths except the cosmic background radiation. For dust opacities in cores, we use the model of OH5 dust since it is successful in reproducing the observed SEDs in star-forming cores (e.g. van der Tak et al. 2000; Evans et al. 2001; Shirley et al. 2002). The inner and outer boundary of the models are set as R$_{in}$ = 80 AU and R$_{out}$ = 8000 AU, which correspond to 0{\farcs}6 and 61{\farcs}5, respectively, and R$_{out}$ is substantically smaller than the separation of the two sources. To compare with models, the observed $I_{\nu}(b)$ are calculated from the 850 {\micron} maps (Shirley et al. 2000), and since RNO91 and L43E can be partially overlapping, the data in sectors that are possibly contaminated were excluded. For example, the data in a 90{\degree} sector at the east pointing toward L43E are cut out in the averaged $I_{\nu}(b)$ of RNO91.

Previous observational studies of pre-protostellar cores (PPCs) suggest a region of roughly constant density at small radii \cite{ward94}, and since they do not have an internal source, the only heating source is the ISRF. We attempt to fit the SED and $I_{\nu}(b)$ of L43E with a set of BE spheres as discussed in Evans et al. (2001). The central density ($n_c$) of the BE sphere is mostly determined by comparing the shape of the normalized $I_{\nu}(b)$. The best-fit values are selected by minimizing $\chi^2$ at 850 {\micron} for $I_{\nu}(b)$ as well as $\chi_{SED}^2$:
$$\chi^2_{850}=\Sigma[\frac{I_{\nu}^{obs}(b_i)-I_{\nu}^{mod}(b_i)}{\sigma_I(b_i)}]^2/N_b,$$
where $I_{\nu}^{obs}(b_i)$ is the normalized intensity at impact parameter $b_i$, $I_\nu^{mod}(b_i)$ is the modeled intensity, $\sigma_I(b_i)$ is the uncertainty of the data, $N_b$ is the number of impact parameters;
$$\chi_{SED}^2=\Sigma[\frac{S_{\nu_i}^{obs}-S_{\nu_i}^{mod}}{\sigma_S(\nu_i)}]^2/N_\nu,$$
where $S_{\nu_i}^{obs}$ is the observed flux, $S_{\nu_i}^{mod}$ is the modeled flux, $\sigma_S(\nu_i)$ is the uncertainty of the data, $N_\nu$ is the number of data points.
 
To compute $\chi_{SED}^2$, the 350 {\micron} flux is excluded because the observing mode of SHARCII used for these observations tends to lose flux for extended sources (Wu et al. 2007). The 160 {\micron} emission is very sensitive to temperature and is used to constrain the ISRF. We tested models with the external A$_V$ in a range between 0.5 to 5.5 and s$_{ISRF}$ between 1 to 10. Figure 7 shows the best-fit model with a BE sphere of $n_c = 3 \times 10^5 cm^{-3}$, s$_{ISRF}$ = 5, and the external A$_V$ = 1. $\chi_{SED}^2$ for the best-fit model is 7.5 (the blue lines in Figure 7), but different combinations of external A$_V$ and s$_{ISRF}$ can provide similar fits, for example, the model with s$_{ISRF}$ = 4 and the external A$_V$ = 0.5 (red lines in Figure 7) and the model with s$_{ISRF}$ = 6 and the external A$_V$ = 1.5 (green lines in Figure 7) have $\chi_{SED}^2$ about 15. The external A$_V$ can be further constrained by line modeling of CO(2-1) emission since photoelectric (PE) heating is sensitive to the surrounding extinction, as detailed in \S 4.2.

For the density distribution of RNO91, we use the gravitational collapse model presented by Shu (1977). In this picture, the model begins from an idealized singular isothermal sphere with an envelope with $n(r)\propto r^{-2}$. Collapse begins inside the sphere, and a wave of infall propagates outward at the speed of sound as time progresses. The density inside the infall radius ({\rinf}) approximates $n(r)\propto r^{-1.5}$ as material falls into the center. For RNO91 at $\lambda \geq 60$ {\micron}, $L_{bol,\lambda \geq 60 {\mu}m} = 0.8$ {\lsun}. The best-fit model adopting this scenario is shown in Figure 7, with an internal heating source of 0.4 {\lsun} and {\rinf} of 0.006 pc.

\subsection{Gas Energetics}
Deep inside the cloud, the gas temperature is equal to the dust temperature because high densities produce rapid collisions between dust and gas. However, at lower densities, since dust and gas are decoupled, the gas temperature must be determined by balancing heating and cooling rates. We use a gas-dust energetics code by S. Doty including gas-dust energy transfer, cosmic ray heating, PE heating, and molecular cooling (details are given in  Doty \& Neufeld [1997] and the appendix in Young et al. [2004]). CO is the major coolant in clouds, and in principle we should include CO depletion in the modeling loop since it could raise $T_K$ in the cold, dense inner region. Goldsmith (2001), Doty \& Neufeld (1997), and Lee (2004) found that the effect is relatively small because of efficient gas-dust coupling at the high densities where CO is significantly depleted; therefore, we did not include depletion and desorption in our gas energetics calculation. 

The calculated $T_K(r)$ is shown in Figure 8. The gas temperature decreases in less dense regions where gas-dust collisions are not as efficient, but rises at the outer edge because of PE heating. In the gas energetics code, the strength of the UV field relative to the local ISRF on the outside of the core is parameterized as G$_0$. We use the equation G$_0$ = exp($-1.8$A$_V$), where the attenuation of the ISRF corresponds to a certain external A$_V$. In dust models, different combinations of external A$_V$ and s$_{ISRF}$ can provide similar fits. For example, Figure 7c shows that models with A$_V$ = 0.4 with s$_{ISRF}$ = 4, A$_V$ = 1 with s$_{ISRF}$ = 5, and A$_V$ = 1.5 with s$_{ISRF}$ = 6 result in similar modeled fluxes at 160 {\micron}. However, the calculated gas temperature profile is more sensitive to the external A$_V$ since the attenuation of the ISRF predominately affects the short wavelengths which are important to the PE heating rate. To examine the model temperature at the outer layer and constrain the amount of surrounding extinction, we modeled the CO(2-1) line toward L43E and compare with the observations (Figure 9), and derive the best-fit G$_0$ of 0.165 (corresponds to an external A$_V$ of 1). Also, the cooling rate depends on the linewidth of CO, and we use a microturbulent width of 0.24 \kms determined by the {\cooo} line fitting. A more detailed treatment of the line modeling is described in \S 4.4.

\subsection{Chemical Models}
We use the evolutionary chemical model presented by Lee et al. (2004). A sequence of BE spheres with increasing central densities and an inside-out collapse initiated at the point of a singular isothermal sphere are combined to describe the dynamics, and the model calculates gas and dust temperatures and follows chemical evolution self-consistently at each time step. The collapse model is adopted from Young \& Evans (2005) for the evolution of luminosity and density distribution. For RNO91, we use the physical parameters that best match the dust modeling results in \S 4.1, and $r_{inf}$ of 0.006 pc corresponds to the timestep of 25000 yr. For L43E, since there is no good constraint for the timescales and dynamics of the PPC stage, we simply assume 8 steps of different central densities and test different timescales for each step to optimize the fitting at  $n_c = 3 \times 10^5 cm^{-3}$. In the chemical models, we assume bare SiO$_2$ for dust grain surfaces to which molecules freeze \cite{bergin95}.

\subsection{Line Modeling}
With the best-fit density profile from dust models, the gas temperature profiles from the gas-energetics calculation, and molecular abundances ($X(r)$) from chemical models, the molecular excitation was modeled with a Monte Carlo radiative transfer code (mc) which runs with 40 shells calculating the level populations. We then simulate the observations with a virtual telescope program (vt) (Choi et al. 1995) at the observed offset position. We use the v$_{LSR}$ derived from observations for the modeled line velocity in each source. To fit the observed lines, we also use simple abundance profiles, such as constant and step function, to compare with the results from chemical models. The goodness of a fit is determined by eye.

We use collision rates by Flower \& Launay (1985) for CO, by Green (1991) for {\form}, by Flower (1999) for {\hcop}, and Daniel et al. (2005) for {\nthp}. For the deuterated species {\dcop}, we use the de-excitation rate coefficients of {\hcop}, and the excitation rate coefficients are obtained by detailed balance. In some cases, rates are linearly extrapolated to lower temperatures. For {\nthp} and {\coooo}, the components of the hyperfine structures are partially resolved. Components separated by less than the 1/e width of the velocity dispersion were aggregated into a single component. The aggregated components are modeled independently with relative intensity-weighted abundances and added together to make combined line profiles.

\section{Modeling Results}
The line modeling results are presented with the observations in Figure 3. The abundance profiles from chemical models are presented in Figure 10. Some of our line models show self-absorption features, which are not seen in the observed line profiles, especially in {\hcop}, H$_2$CO, and {\nthp}. An explanation for the lack of self-absorption in molecular lines is clumpy cloud structure with the existence of macroturbulent motions, which is simulated with the Monte Carlo methods for 3-dimensional clouds by Juvela (1997). The line profiles can be confused with heavy self-absorption as well as rotation (e.g. Pavlyuchenkov et al. 2008). Since we are not including a complete 3-dimensional calculation in this paper, we only fit the line strength at half maximum level for lines that are self-absorbed in the models.

From observations toward star forming cores, one of the features that potentially indicates the evolutionary status is an anti-correlation of CO and {\nthp}. {\nthp} is thought to be a good tracer of dense core gas at earlier evolutionary stages until $T_D$ reaches the temperature for CO evaporation. The relative depletion and desorption of N$_2$ and CO are the key processes governing the abundance of {\nthp}; however, recent laboratory result shows there is no significant difference of N$_2$ and CO binding energies (\"{O}berg et al. 2005; Bisschop et al. 2006). With an updated higher value of N$_2$ binding energies, the modeled {\nthp} lines are much weaker than the observed lines. The timescale for the PPC stage is also a key issue for the freezeout of molecules, so we varied the timescales at different central densities. A longer total timescale allows more N$_2$ to form, but a longer time at higher densities ($n > 10^5 cm^{-3}$) would result in more depletion. {\nthp} abundances are directly affected by the amount of N$_2$ formation and CO depletion. To fit the observed {\nthp} lines, we assume a longer total timescale that allows more N$_2$ formation and CO depletion and shorter time at later stages to prevent N$_2$ depletion. Two test models are listed in Table 4. We use the timescales in Lee et al. (2004) as the standard model (Model 1), and to optimize the {\nthp} abundances for L43E, a longer total timescale of $2\times10^6$ yr is assumed (Model 2). In Figure 10, the solid lines are abundance profiles from Model 1, and the dashed lines are from Model 2. For both models, we also tested the {\nthp} dissociative recombination rates in Geppert et al. (2004) to compare with the rates in the UMIST database for astrochemistry 1999 \cite{leteuff00}. The rates in Geppert et al. (2004) fit the observations better, and are included in our chemical network for the models presented in this paper.

The CO lines are difficult to fit in detail since they are very optically thick and confused by the wing components. The line profile toward RNO91 has wide outflow wings, and the line profile toward L43E also shows a blue wing which possibly comes from another component in the foreground along the line of sight. The modeled CO abundances can still be tested through line fitting of CO isotopologues. We assume the abundance profiles for the CO isotopologues have the same form as CO with an isotope ratio of O/$^{18}$O $=580$ (Wilson \& Rood 1994) and $^{18}$O/$^{17}$O $=4$ \cite{wouterloot05}. In Model 2, the modeled {\cooo} and {\coooo} line intensities are efficiently decreased to fit the data since more CO depletion is produced. We also assume an ortho to para ratio of 1.5 (Dickens \& Irvine 1999) to model the ortho-{\form} ($3_{1,2}-2_{1,1}$) lines. For L43E, the {\form} abundance in Model 2 is higher than in Model 1 by a factor of 2, but it is possible that self-absorption has greatly weakened the blue peak.

For the {\hcop} lines toward L43E, although the observed line profile might be obscured by absorption from a foreground component, the modeled line is still too strong. The cosmic ray ionization rates could affect the results, but neither enhanced nor lowered reaction rates improved the fit of both {\hcop} and {\nthp} data in both cores simultaneously. We tested empirical abundance profiles with lower abundance at large radii. For example, if G$_0$ is higher than we adopted, the higher dissociative recombination rate can lower the abundance in the outer region. However, we did not find an improved fit even with extremely low {\hcop} abundances in the outer region. To match the line intensity according to the red peak, we use a constant abundance $X$({\hcop}) = $4 \times 10^{-10}$, which is lower than the chemical model prediction by a factor of $\sim$7.5. For the deuterated species, since they are favored in a cold region with CO depletion, both {\dcop} and {\ntdp} are enhanced relative to {\hcop} and {\nthp} at high densities. The {\dcop} abundances in L43E are increased from Model 1 to Model 2 because of the amount of CO depletion, but the modeled {\dcop} lines from chemical models are still weaker than observed. The {\dcop} line emission generated primarily in the center region with high densities where {\hcop} can be depleted. It is possible that there is still more CO depletion than the chemical models predict. A {\dcop} abundance increased by a factor of $\sim$8 is needed to fit the observed line intensity for L43E, and for RNO91, we use a step function with depletion factor of 5 within 0.003 pc. The best-fit empirical abundance profiles are shown in Figure 10 with blue dotted lines for L43E and red dotted lines for RNO91. To compare with the results in \S 3.3.3, we also calculated the optical depths for the best-fit {\hcop} lines with empirical abundances. The optical depth of the simulated {\hcop}(3-2) line in RNO91 is 23, which is much higher than the value of 3 in L43E. This result also supports the analysis in \S 3.3.3 that the $N$({\dcop})/$N$(\hcop) factors calculated with optically thin approximation provide a lower limit, with $N$({\dcop})/$N$(\hcop) at least 4.8 times more enhanced in L43E than in RNO91.


To summarize, most of the lines in both sources can be fitted reasonably well with Model 2, which has a longer timescale for the PPC stage. However, the change of parameters in our different chemical models may only be compensating the effects not yet included. For example, the grain properties should also be time-dependent. Grain mantles covered by water-rich ice have strong binding energies for molecules to freeze on, but N$_2$ forms slowly and is less affected. As the density increases and CO freezes out, the grain mantle surfaces can be contaminated by CO ice. Once grain mantles are CO-dominant, the binding energies are lower, which results in less depletion of N$_2$ and {\nthp} compared with {\hcop}, since {\hcop} depletion follows CO depletion. Our modeling results of longer timescales at low densities may work equivalently to the effects of higher grain surface binding energies (water-dominant) at lower densities. The difference between our modeling results and observations may be more specifically explained in a model including time-dependent grain properties. Heavy CO (and thus {\hcop}) depletion on the initial water-dominant grain surfaces can result in more abundant {\dcop}. With time, grain surfaces are covered with CO (which produces CO$_2$ on surface), resulting in less depletion of N$_2$ and more abundant {\nthp}.    


\section{Summary}
We have presented Spitzer observations and molecular line observations from the CSO toward the two sources RNO91 and L43E evolving in the same region. A self-consistent modeling procedure is employed to investigate the physical condition in each source, and to compare the chemistry in an evolutionary picture. Our results are as follows.

From the Spitzer images, RNO91 can be categorized as Class I, and L43E is still starless with no embedded source detected. CO depletion is seen in both sources, but the degree of depletion is higher in L43E. Also, strong {\dcop} enhancement is seen in L43E.

Dust modeling results showed that L43E can be fitted by a BE sphere with central density of $3\times10^5 cm^{-3}$, and RNO91 is characterized to have $r_{inf}$ of 0.006 pc with an inside-out collapse model (Shu 1977). For the line modeling, the best-fitting chemical model suggests a longer total timescale at earlier stages but fast evolution at higher densities for the PPC stage. The observed abundances of {\nthp} and CO isotopologues can be matched well in both sources, but in L43E, the modeled {\hcop} abundance is too high by a factor of 7.5, and the modeled {\dcop} abundance is too low by a factor of 8. It is possible there is still more CO depletion in the center than the model predicted. In this picture, RNO91 is possibly the first star forming in this region, and L43E could be spending a longer time at the PPC phase. 

\acknowledgements
We would like to thank Yancy Shirley for providing the radial intensity profiles from the SCUBA data, Jingwen Wu, Michael Dunham, Miranda Nordhaus, and Hyo Jeong Kim for their help to obtain the data at the CSO, and Ted Bergin for the chemical code. Support for this work, part of the Spitzer Legacy Science Program, was provided by NASA through contract 1224608 issued by the Jet Propulsion Laboratory, California Institute of Technology, under NASA contract 1407. This work has also been supported by NSF Grants AST-0307250, AST-0607793, and NASA Origins grant NNX07AJ72G. Jeong-Eun Lee gratefully acknowledges the support by the Korea Science and Engineering Foundation (KOSEF) under a cooperative agreement with the Astrophysical Research Canter for the Structure and Evolution of the Cosmos (ARCSEC).


\begin{figure}[htp]
\begin{center}
\includegraphics[width=0.8\textwidth]{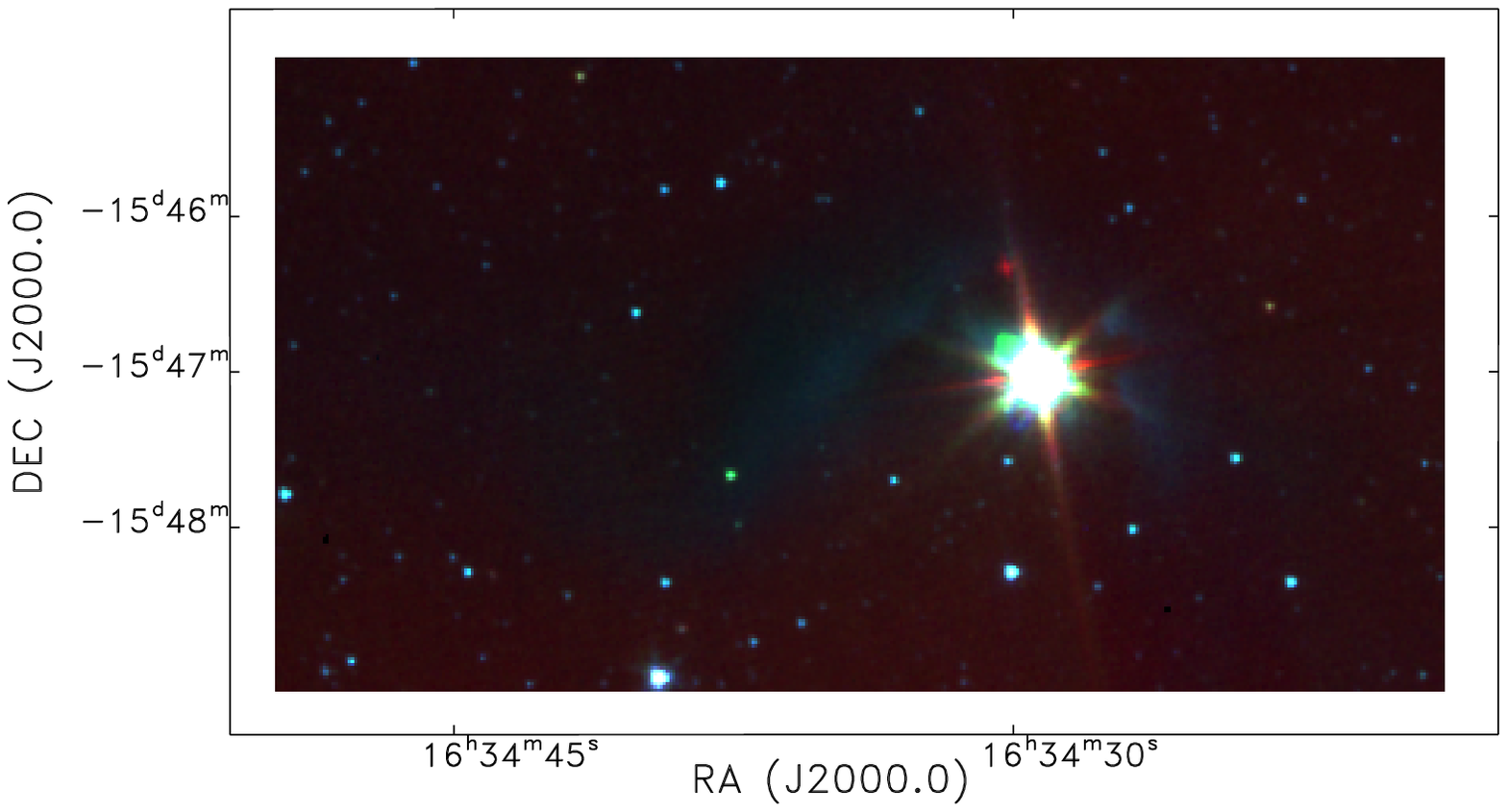}
\caption{Three color image composite of IRAC1 (blue), IRAC2 (green); IRAC4 (red) toward L43. The bright object is RNO91, and L43E is the dark region in the east.}
\end{center}
\end{figure}

\begin{figure}[htp]
\begin{center}
\includegraphics[width=0.8\textwidth]{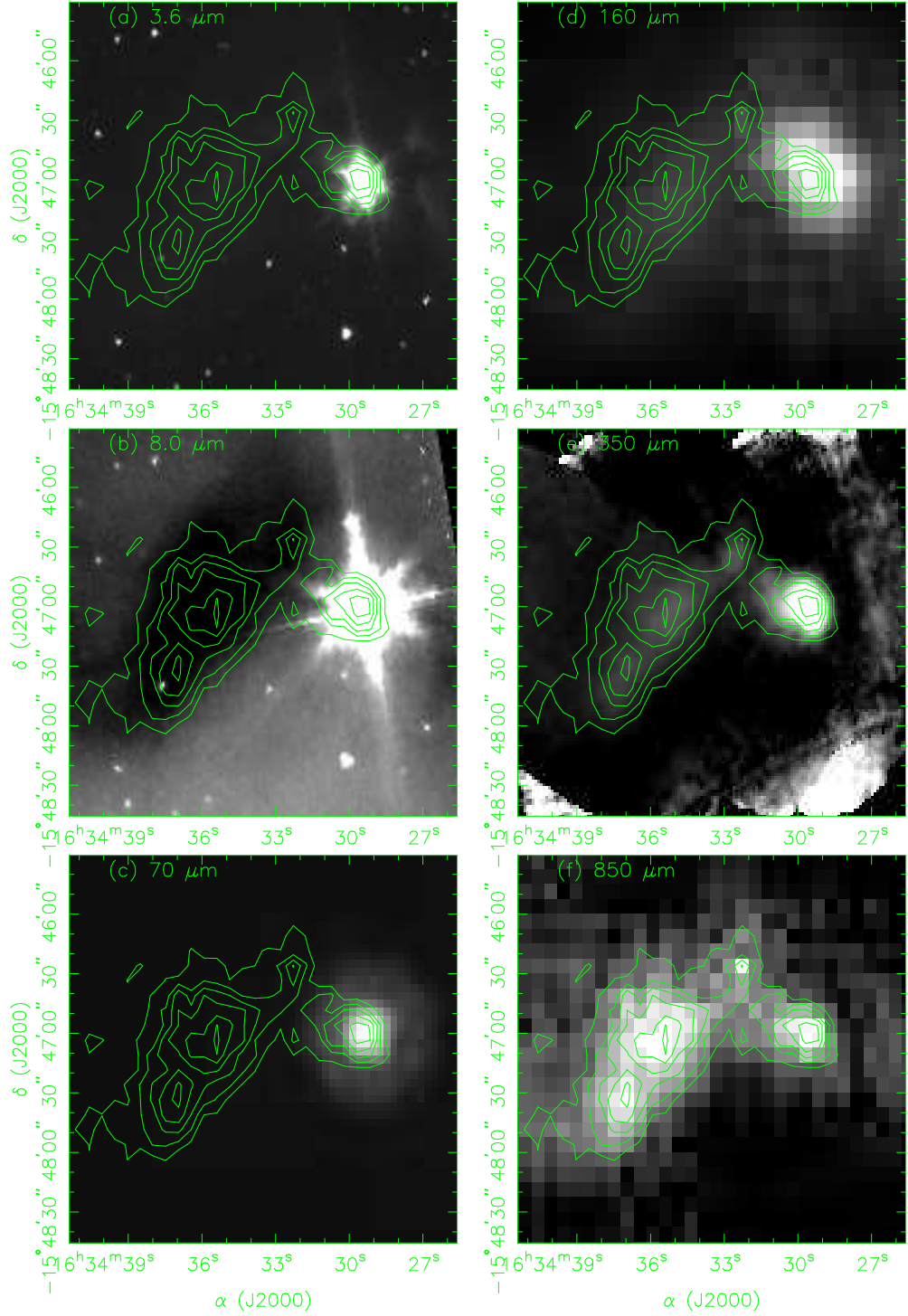}
\caption{SCUBA 850 {\micron} contours on images at (a) IRAC 3.6 {\micron} from c2d, (b) IRAC 8.0 {\micron} from Spitzer Program 20386, (c) MIPS 70 from Spitzer Program 30384 (d) MIPS 160 {\micron} from Spitzer Program 30384, (e) SHARCII 350 {\micron}, and (f) SCUBA 850 {\micron}.}
\end{center}
\end{figure}

\begin{figure}[htp]
\begin{center}
\includegraphics[width=0.8\textwidth]{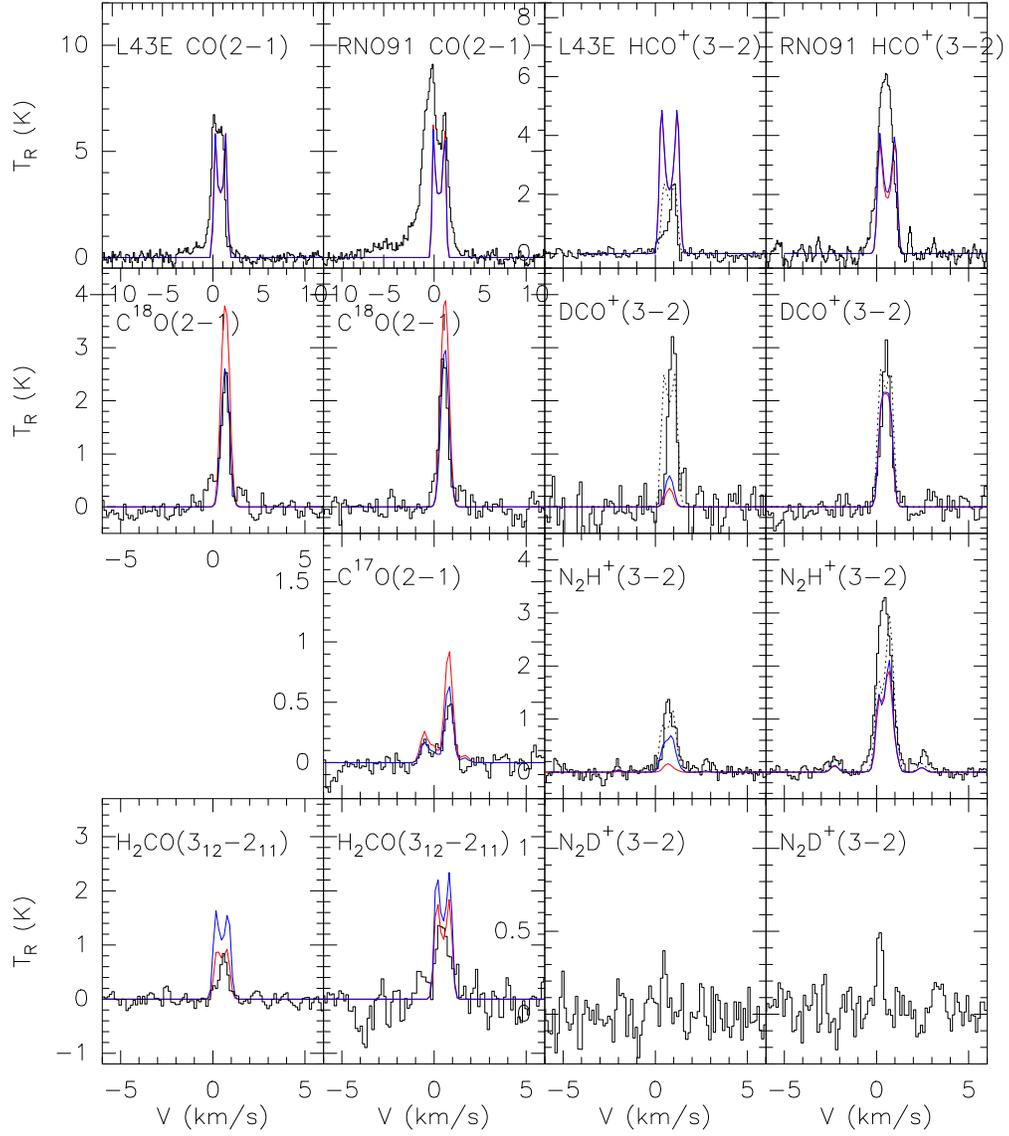}
\caption{Molecular line observations toward RNO91 and L43E, shown as black solid histograms, and the modeled lines from Model 1 (red lines), Model 2 (blue lines), and empirical models for {\hcop}, {\dcop}, and {\nthp} (black dotted lines). Model 2 has a longer timescale for the PPC stage.} 
\end{center}
\end{figure}
\clearpage

\begin{figure}[htp]
\begin{center}
\includegraphics[angle=270,width=0.8\textwidth]{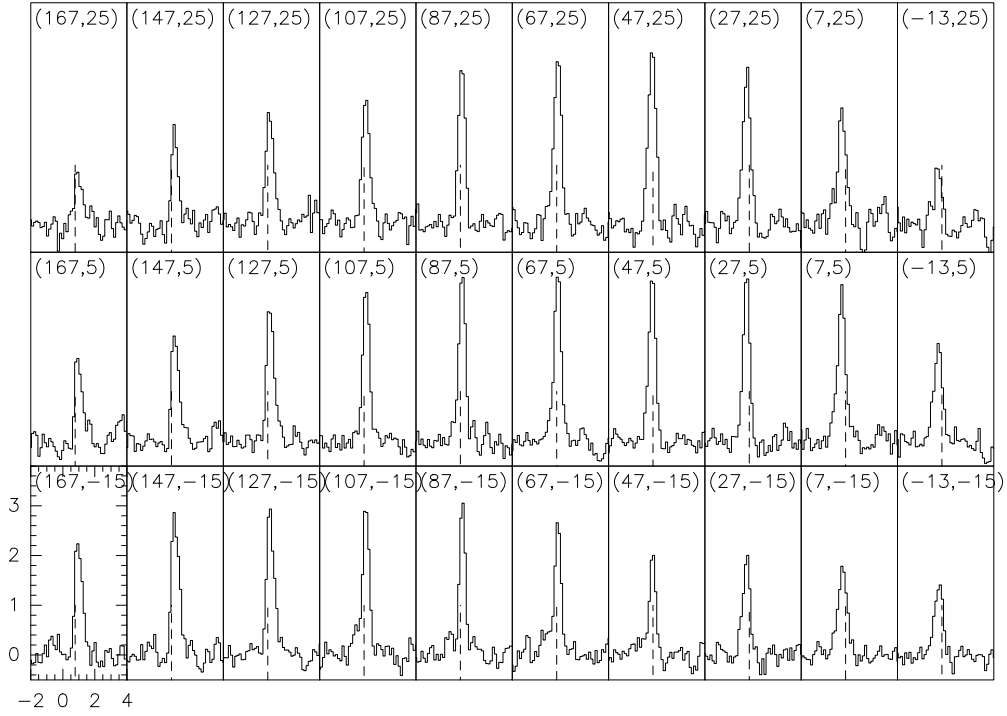}
\caption{Map of {\dcop}(3-2) in the L43 region. The spectra are convolved to the position of RNO91 at (7{\as},5{\as}) in 20{\as} steps. The dashed lines indicate v$_{LSR}$ = 0.77 {\kms} for L43E at (89{\as},6{\as}). The velocities toward the east are red-shifted relative to L43E, and the velocities toward the west are blue-shifted.}
\end{center}
\end{figure}

\begin{figure}[htp]
 \begin{center}
 \includegraphics[width=0.8\textwidth]{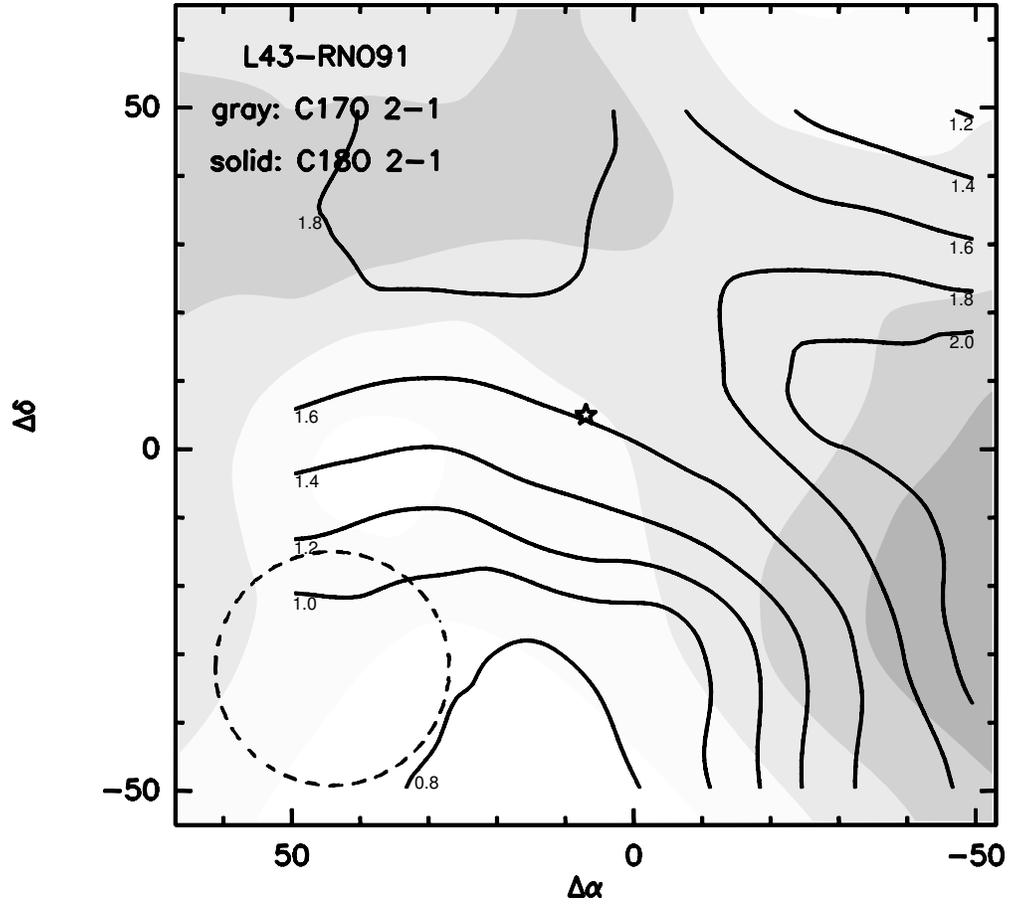}
 \caption{Integrated intensity map of {\cooo}(2-1) (solid lines) and {\coooo}(2-1) (grayscale) convolved in 30{\as} steps. The {\cooo} contour levels start at 0.8 {\kms} and increase by 0.2 {\kms}. The {\coooo} contour levels start at 3$\sigma$ and increase by 2$\sigma$. The star marks the position of RNO91, and there is no obvious intensity peak in the mapped region in either line. The {\coooo}(2-1) beam is shown in dashed line.}
 \end{center}
 \end{figure}

\begin{figure}[htp]
 \begin{center}
\includegraphics[width=0.8\textwidth]{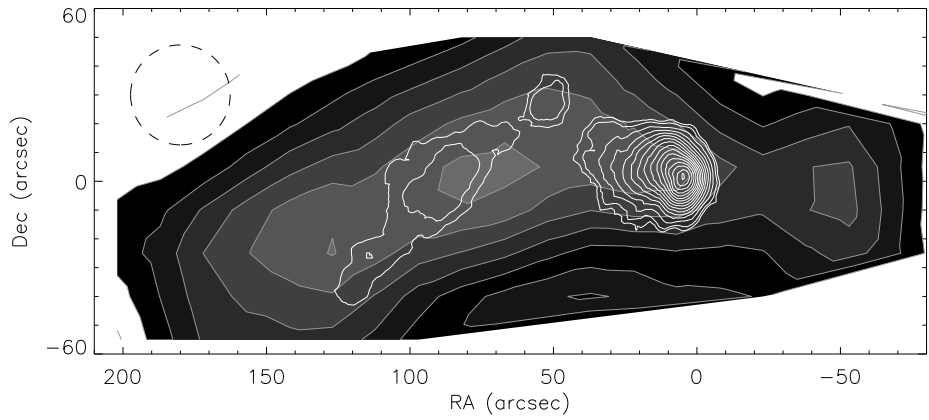}
 \caption{SHARCII 350{\micron} contours (solid white) overlaid with {\dcop}(3-2) integrated intensity map (grayscale). The 350{\micron} coutour levels are 2$\sigma$, 3$\sigma$, 5$\sigma$, and increased by 2$\sigma$. The {\dcop} integrated intensity contours start at 3$\sigma$ and increase by 2$\sigma$. The {\dcop}(3-2) beam is shown in dashed lines.}
 \end{center}
 \end{figure}

\begin{figure}[htp]
 \begin{center}
\includegraphics[width=0.8\textwidth]{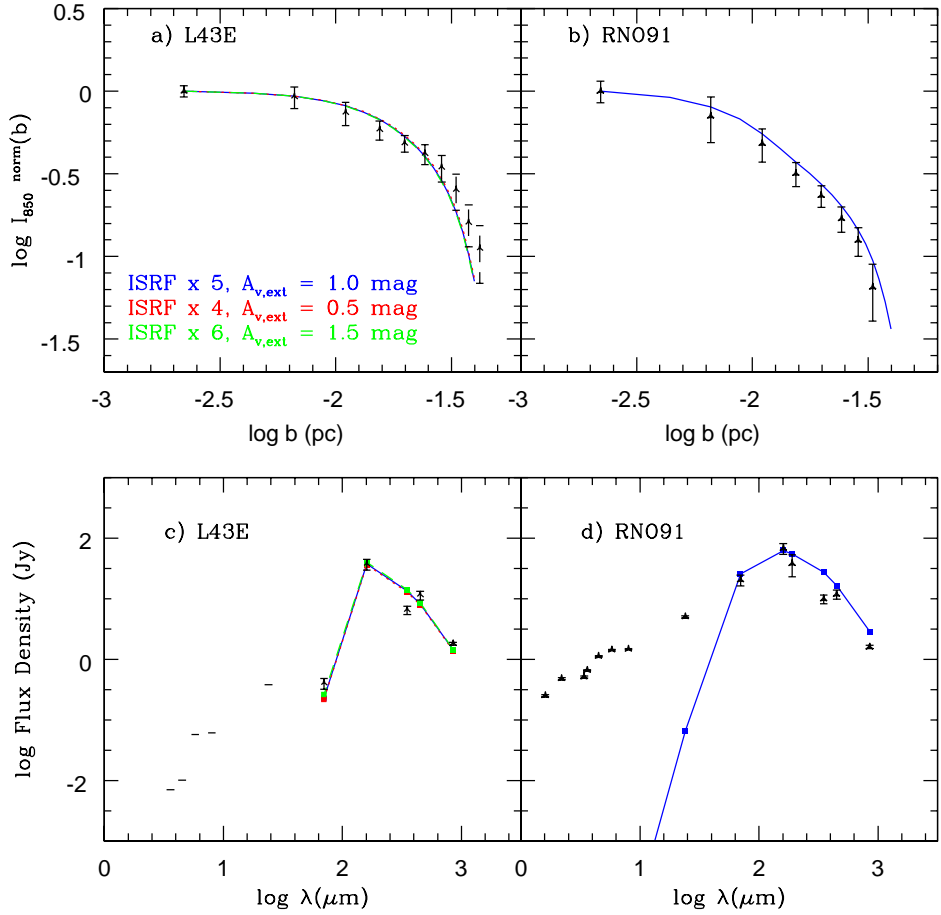}
 \caption{Best-fit dust models for both RNO91 and L43E. a) and b): The model fitting of the normalized $I_{850}(b)$; c) and d): the SED fitting. The solid blue lines are the model with an external A$_V$ = 1 and $s_{ISRF}=5$, the dotted red lines are the model with an external A$_V$ = 0.5 and $s_{ISRF}=4$, and the dashed green lines are the model with an external A$_V$ = 1.5 and $s_{ISRF}=6$. The different combinations provide similar fits and the modeled lines are on top of each other. The short horizontal lines are marks for the upper limits for L43E from Spitzer Observations. 
}
 \end{center}
 \end{figure}

 \begin{figure}[htp]
 \begin{center}
 \includegraphics[width=0.8\textwidth]{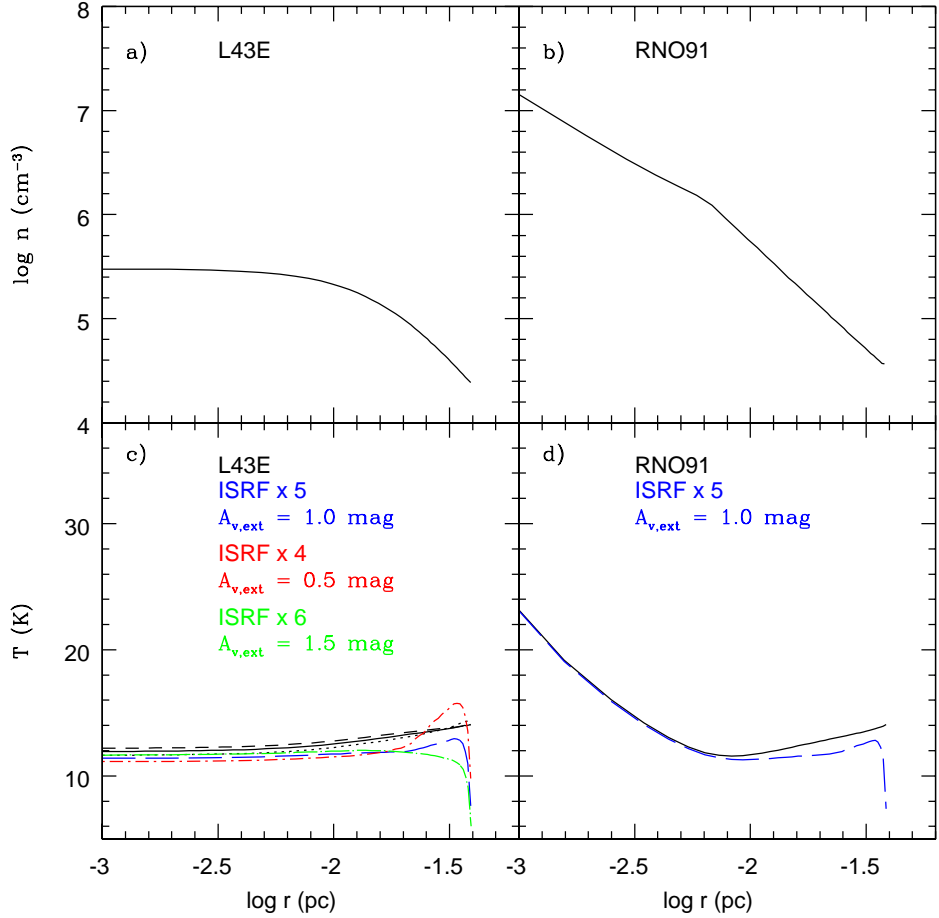}
 \caption{Density profiles and calculated $T_D$ and $T_K$ with the same models in Figure 5. a): The density profile of BE sphere with $n_c=3\times10^5 cm^{-3}$; b): the density profile of inside-out collapse model with $r_{inf}=0.006$ pc; c) and d): calculated $T_D$ and $T_K$.  The solid black lines are $T_D(r)$ from the model with $s_{ISRF}=5$ and A$_V$ = 1, the dashed black lines are $T_D(r)$ from the model with $s_{ISRF}=4$ and A$_V$ = 0.5, the dotted black lines are $T_D(r)$ from the model with $s_{ISRF}=6$ and A$_V$ = 1.5, the long-dashed blue lines are $T_K(r)$ from the model with $s_{ISRF}=5$ and A$_V$ = 1, the dot-short-dashed lines are $T_K(r)$ from the model with $s_{ISRF}=4$ and A$_V$ = 0.5, and the dot-long-dashed green lines are $T_K(r)$ from the model with $s_{ISRF}=4$ and A$_V$ = 1.5}
 \end{center}
 \end{figure}

\begin{figure}[htp]
\begin{center}
\includegraphics[width=0.8\textwidth]{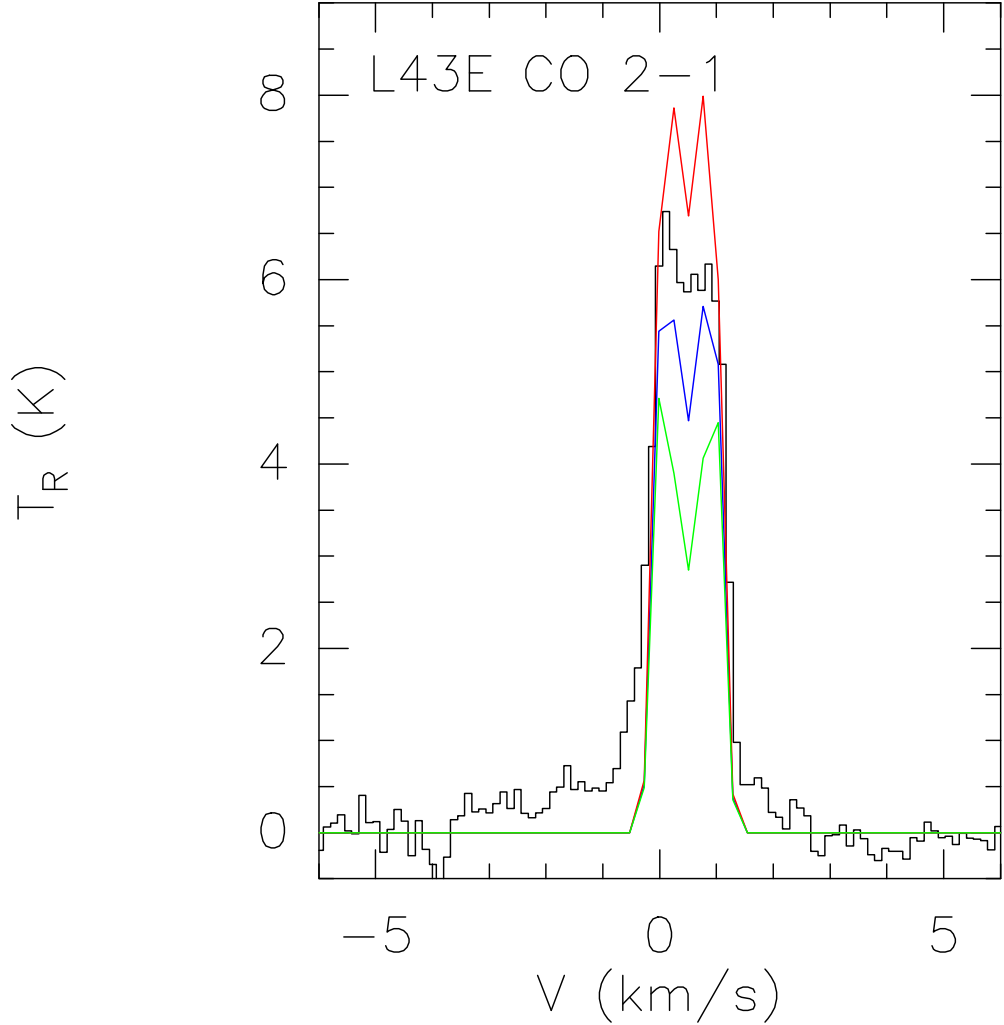}
\caption{Simple CO line models with the results in figure 7. The blue line is from the model with $s_{ISRF}=5$ and A$_V$ = 1, the red line is from the model with $s_{ISRF}=4$ and A$_V$ = 0.5, and the green line is from the model with $s_{ISRF}=4$ and A$_V$ = 1.5. X(CO) is assumed $3\times10^{-5}$ in the models, but a higher or lower value only affects the size of the self-absorption dip. The modeled line velocity is shifted to 0.5 {\kms} to be aligned with the observed line. The blue wing in the observed CO line profile might come from another component in the cloud.}
\end{center}
\end{figure}
\clearpage

\begin{figure}[htp]
\begin{center}
\includegraphics[width=0.8\textwidth]{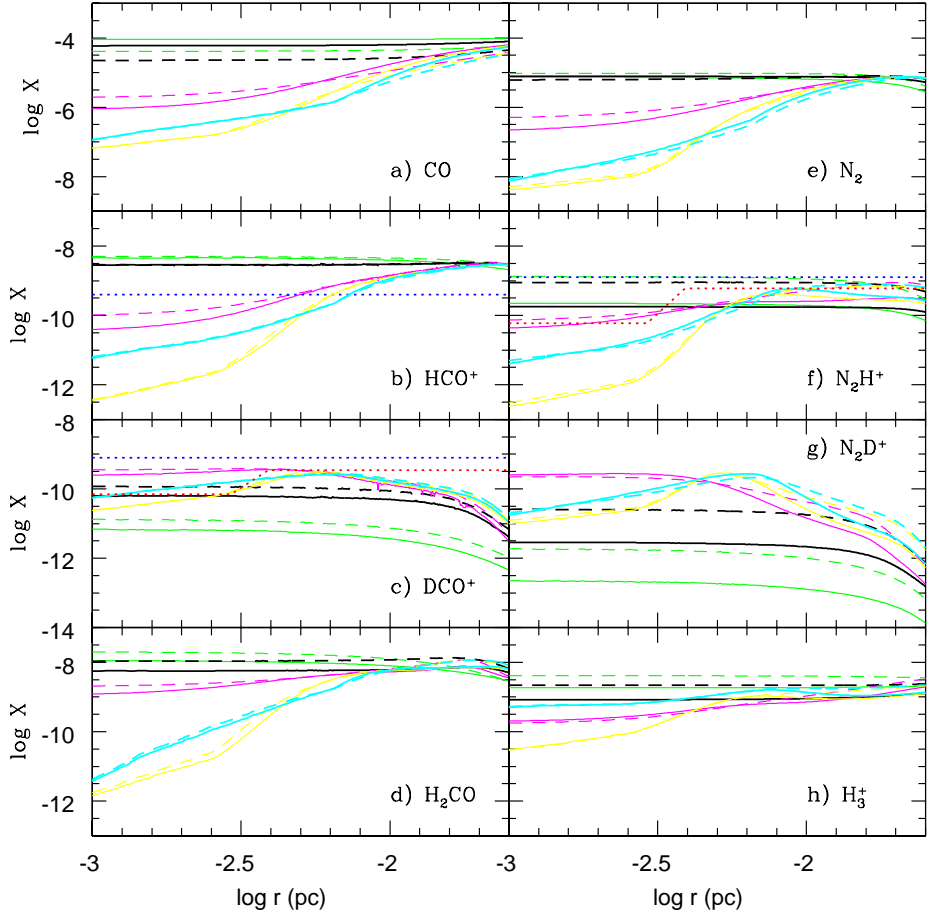}
\caption{Evolution of abundance profiles for a) CO, b) {\hcop}, c) {\dcop}, d) {\form}, e) N$_2$, f) {\nthp}, g) {\ntdp}, and h) H$_3^+$ at selected timesteps of Model 1 (solid lines) and Model 2 (dashed lines). The difference between the two models is that Model 2 has a longer PPC timescale. For t $<$ 0 (before collapse), the lines shows BE spheres of central densities $10^5$ (green), $3\times10^5$ (black), and $10^7$ (magenta) cm$^{-3}$. The central densities correspond to different timesteps in each model, and the timesteps are listed in Table 4. For t $>$ 0 (after collapse), the timesteps are $t=10^4$ (yellow) and $t=2.5\times10^5$ (cyan) yr. The blue dotted lines are best-fit empirical models for {\hcop}, {\dcop}, and {\nthp} in L43E. The red dotted lines are best-fit empirical models for {\dcop} and {\nthp} in RNO91.}
\end{center}
\end{figure}
\clearpage


\begin{table}[htp]
\caption{CSO Observing Log}
\small
\begin{tabular}{lccccc}
\tableline
Line & Frequency & UT Date & Beam Width & $\eta_{MB}$ & $\delta$v \\
 & (MHz) & & (arcsec) & & (km $s^{-1}$) \\\tableline
$^{12}$CO 2 -- 1 & 230537.970 & 2005 Mar & 33 & 0.82 & 0.10 \\
 & & 2006 Jun & 32 & 0.77 & 0.13 \\
{\coooo} 2 -- 1 &   224714.400 & 2000 Jun & 33 & 0.81 & 0.17 \\
& 224714.368\tablenotemark{a} & 2006 Jun & 33 & 0.77 & 0.13 \\ 
{\cooo} 2 -- 1 & 219560.352 & 1998 Jul & 28 & 0.57 & 0.15 \\
 	&	& 2005 Mar & 34 & 0.82 & 0.11 \\
 	&	& 2005 Jun & 34 & 0.74 & 0.11 \\
& & 2006 Jun & 34 & 0.77 & 0.13 \\
CS 7 -- 6 & 342882.900 & 2000 Jun & 22 & 0.77 & 0.10 \\
{\form} 3$_{12}$ -- $2_{11}$ & 225697.787 & 2002 Jun & 33 & 0.59 & 0.15 \\
 & & 2005 Jun & 33 & 0.74 & 0.13\\ 
{\hcop} 3 -- 2 & 267557.620 & 2000 Jun & 28 & 0.81 & 0.14\\
 & & 2006 Jun & 28 & 0.77 & 0.13 \\
{\dcop} 3 -- 2 & 216112.605 & 1999 Jul & 35 & 0.60 & 0.15 \\
& & 2006 Jun & 35 & 0.77 & 0.14 \\ 
{\nthp} 3 --2  & 279511.701 & 2003 Jul & 27 & 0.68 & 0.14\\
 {\ntdp} 3 -- 2 & 231321.635 & 2003 Jul & 32 & 0.70	& 0.13\\
 & & 2005 Jun & 32 & 0.74 & 0.11\\\tableline	
\end{tabular}
\tablenotetext{a}{Reference frequency for the hyperfine shifts in Ladd et al. (1998)}
\end{table}
\clearpage

\begin{table}[htp]
\caption{Observed Flux Densities}
\small
\begin{tabular}{llllc}
\tableline
Source & $\lambda$ & $S_\nu$($\sigma$) & Aperture & Reference \\
 & ({\micron}) & (mJy) & (arcsec) & \\\tableline
RNO91 &	1.6    & 250(10)      & 12 & 1\\
      &	1.65   & 143(6.86)    & 2.5\tablenotemark{a} & 2\\
      & 2.17   & 274(8.34)    & 2.5\tablenotemark{a} & 2\\
      &	2.2    & 480(20)      & 12 & 1\\
      &	3.4    & 510(20)      & 12 & 1\\
      &	3.6    & 737(51.1)    & 1.7\tablenotemark{b} & 3\\
      & 4.5    & 1110(66.9)   & 1.7\tablenotemark{b} & 3\\
      & 5.8    & 1440(72.0)   & 1.9\tablenotemark{b} & 3\\
      & 8.0    & 1660(82.8)   & 2.0\tablenotemark{b} & 3\\
      & 24     & 5000(200)\tablenotemark{c}    & 70 & 4\\ 
      & 70     & 20300(4060)\tablenotemark{c}  & 70 & 4\\      
      &	160    & 67400(13480)\tablenotemark{c} & 96 & 4\\
      & 190    & 38000(15200) & 60 & 5\\
      &	350    & 9900(1600)   & 40 & 6\\
      &	450    & 11800(2000)  & 40 & 7\\
      &	850    & 1600(90)     & 40 & 7\\\tableline
L43E  & 3.6    & $<$ 6.99\tablenotemark{d}  & 1.7\tablenotemark{b} & 3\\
      & 4.5    & $<$ 10.17\tablenotemark{d} & 1.7\tablenotemark{b} & 3\\
      & 5.8    & $<$ 57.3\tablenotemark{d}  & 1.9\tablenotemark{b} & 3\\
      & 8.0    & $<$ 61.5\tablenotemark{d}  & 2.0\tablenotemark{b} & 3\\
      & 24     & $<$ 378\tablenotemark{d}   & 5.7\tablenotemark{b} & 3\\
      & 70     & 400(80)\tablenotemark{c} & 70 & 4\\
      & 160    & 37000(7400)\tablenotemark{c} & 128 & 4\\
      &	350    & 6500(1300)  & 40 & 6\\
      & 450    & 11400(1900) & 40 & 7\\
      & 850    & 1800(100)   & 40 & 7\\\tableline
\end{tabular}
\tablenotetext{a}{FWHM of 2 Micron All Sky Survey point-spread profile}
\tablenotetext{b}{FWHM of Spitzer point-spread profile}
\tablenotetext{c}{Uncertainty given in c2d delivery documentation (Table 21)}
\tablenotetext{d}{3-$\sigma$ value estimated with the Sensitivity Performance Estimation Tool (SENS-PET) from Spitzer Science Center}
\tablenotetext{}{REFERENCES.--- (1) Myers et al. 1987; (2) 2 Micron All Sky Survey; (3) results from Spitzer c2d; (4) results from Spitzer Program 30384; (5) Ladd et al. 1991; (6) Wu et al. 2007; (7) Shirley et al. 2000}
\end{table}
\clearpage

\begin{table}[htp]
\caption{CSO Observational Results}  
\small
\begin{tabular}{lllllllll}
\tableline
Source & Line & I($\sigma$) & $v_{LSR}$ & $\Delta$v & $T_R$ & offset\tablenotemark{a} & N$_{obs}$\tablenotemark{b}\\
     & &(K km $s^{-1}$) & (km $s^{-1}$) & (km $s^{-1}$) & (K) & (arcsec) & \\\tableline
RNO91 & $^{12}$CO 2 -- 1\tablenotemark{c} &  22.50(0.19)&   0.78(0.10)	& 2.69(0.08)    &  7.87(0.21) & (6, 3) & 1\\
&{\coooo} 2 -- 1\tablenotemark{d}  & 0.39(0.08)&   0.57(0.02)  & 0.48(0.06)  & 0.49(0.05) & (7, 5) & 169\\
&{\cooo} 2 -- 1 &  1.71(0.06)  & 0.51(0.01) & 0.59(0.03)&  2.74(0.16) & synthesized to (7, 5) & 121\\
&CS 7 -- 6	&       -	&    -		& -	     & $<$0.40 (r.m.s 0.11) & (6, 3) & 1\\
&{\form} 3$_{12}$ -- $2_{11}$ & 1.47(0.19)  & 0.66(0.06) & 1.14(0.21) &  1.21(0.27) & (6, 3) & 1\\
&{\hcop} 3 -- 2 & 5.99(0.08)  & 0.48(0.01) & 0.89(0.01) &  6.33(0.21) & (0, 0) & 1\\
&{\dcop} 3 -- 2 & 1.85(0.07)  & 0.52(0.01)  & 0.59(0.03)&  2.92(0.16) & synthesized to (7, 5) & 342\\
&{\nthp} 3 --2\tablenotemark{c} & 3.78(0.07) &  0.40(0.01) & 0.64(0.03)&   3.29(0.07) & (7, 5) & 11\\
& {\ntdp} 3 -- 2 & 0.21(0.03)&   0.18(0.03) & 0.37(0.06) &  0.53(0.01)	& (7, 5) & 5\\\tableline
L43E & $^{12}$CO 2 -- 1&  10.10(0.13)& 0.46(0.01) &  1.38(0.02)& 6.88(0.15) & (116, -13) & 5\\
&{\cooo} 2 -- 1 &  1.40(0.09) & 0.66(0.02) & 0.55(0.05)& 2.38(0.18) & (96, -13) & 6\\
&{\form} 3$_{12}$ -- $2_{11}$ & 0.47(0.04)  & 0.59(0.03) & 0.58(0.07)& 0.76(0.01) & (89, 6) &1\\
&{\hcop} 3 -- 2 & 1.11(0.07) & 0.94(0.01)  & 0.47(0.04) & 2.24(0.10) & (89, 6) &5\\
&{\dcop} 3 -- 2 & 1.65(0.05) & 0.88(0.01)  & 0.50(0.02) & 3.13(0.10) & (90, 10) &1\\
&{\nthp} 3 -- 2\tablenotemark{c} & 1.19(0.10) &  0.73(0.01)& 0.23(0.03)&  1.36(0.10) & (86, 3) &25\\
& {\ntdp} 3 -- 2 & 0.11(0.03) & 0.48(0.03) & 0.26(0.07) & 0.39(0.11) & (96, -13) &5\\\tableline
\end{tabular}
\tablenotetext{a}{(0,0) is the reference position of (16$^h$34$^m$29.0$^s$, $-$15\degree47\am3.2\as) (J2000.0), and the centroids of RNO91 and L43E are at (7, 5) and (89, 6), respectively.}
\tablenotetext{b}{Number of observed positions.}
\tablenotetext{c}{Double-peaked line: $T_R$ refers to the stronger peak, v$_{LSR}$ refers to the velocity of the dip, I is of all area, and $\Delta$v is I divided by $1.06T_R$ for the gaussian equivalent linewidth.}
\tablenotetext{d}{Hyperfine structure: v$_{LSR}$ and $\Delta$v are from hyperfine structure fitting, I is of all area, and $T_R$ refers the main peak.}
\end{table}
\clearpage

\begin{table}[htp]
\caption{Chemical Models} 
\small 
\begin{tabular}{rrr}
\tableline
& Model 1 & Model 2\\\tableline
Central Density & Timestep & Timestep\\
(cm$^{-3}$) & (yr) & (yr)\\\tableline
10$^7$ & $5\times10^3$ & $3\times10^3$\\
$3\times10^6$ & $1\times10^4$ & $5\times10^3$\\
10$^6$ & $2\times10^4$ & $1.2\times10^4$\\
$3\times10^5$ & $3\times10^4$ & $3\times10^4$\\
10$^5$ & $6\times10^4$ & $10^5$ \\
$3\times10^4$ & $1.25\times10^5$ & $3.5\times10^5$ \\
10$^4$ & $2.5\times10^5$ & $1.5\times10^6$ \\
10$^3$\tablenotemark{a} & $2.5\times10^5$ & $1.5\times10^6$
\end{tabular}
\tablenotetext{a}{The initial constant density.}
\end{table}
\clearpage

\end{document}